\documentclass{epl}
\usepackage{amsmath}

\title{Absence of Landau's Diamagnetism in Two Dimensions}
\author{S. Fujita\inst{1}\thanks{E-mail: \email{fujita@buffalo.edu}}
\and H. C. Ho\inst{2}\thanks{E-mail: \email{hcho@phys.cts.nthu.edu.tw}}}

\institute{
  \inst{1} Department of Physics, University at Buffalo, SUNY\\
           Buffalo, New York 14260, USA\\
  \inst{2} Physics Division, National Center for Theoretical Sciences\\
           Hsinchu 30013, Taiwan
}


\begin{document}

\maketitle

\begin{abstract}
The statistical weight $W$ as a function of energy $E$ for
quasielectrons with mass $m^{\ast}$ subject to a fixed magnetic field
$\mathbf{B}$ is $W/A=\left(m^{\ast}/\pi\hbar^{2}\right)E+\left(eB/\pi^{2}\hbar\right)\sum_{\nu=1}^{\infty}(-1)^{\nu}\nu^{-1}\cdot\sin\left(2\pi\nu E/\hbar\omega_\text{c}\right)$,
where $A$ is the sample area, and $\omega_\text{c}\equiv eB/m^{\ast}$ the
cyclotron frequency. Significantly, there is no Landau's term proportional
to $B^{2}$ in 2D. This leads to the conclusion that the 2D electron
system is always paramagnetic, but shows a magnetic oscillation.
\end{abstract}

Landau~\cite{Landau1930}, Sondheimer and Wilson~\cite{Wilson1954,Sondheimer1951}
discussed the de Haas-van Alphen (dHvA) oscillation~\cite{deHaas1930}
of a three-dimensional (3D) system of quasifree electrons. Nakamura~\cite{Nakamura1965}
calculated the statistical weight $W$ associated with the Landau
states, and treated the dHvA oscillation. We extend Nakamura's theory
to a 2D system.

Let us consider a dilute system of electrons, each with effective mass
$m^{\ast}$, moving in a plane. Applying a magnetic field $\mathbf{B}$
perpendicular to the plane, each electron will be in a Landau state
of energy%
\begin{equation}
\label{eq:E,NsubL}
E=\left(N_\text{L}+1/2\right)\hbar\omega_\text{c},\quad
N_\text{L}=0,1,\cdots,
\end{equation}%
where $\omega_\text{c}=eB/m^{\ast}$ is the cyclotron frequency. The
degeneracy of the \emph{Landau level} (LL) is%
\begin{equation}
\label{eq:eBAover2pihbar,A}
\frac{eBA}{2\pi\hbar},\quad A=\textrm{sample area}.
\end{equation}

\begin{figure}
\onefigure{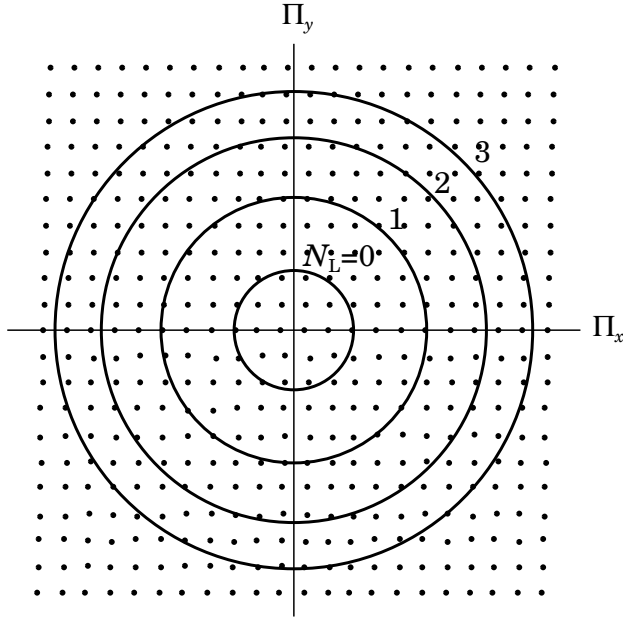}
\caption{Quantization scheme for free electrons in 2D: without a magnetic field (dots),
and in a magnetic field (circles).}
\label{cap:Landau_Quant}
\end{figure}

 We introduce kinetic momenta%
\[
\Pi_{x}=p_{x}+eA_{x},\quad\Pi_{y}=p_{y}+eA_{y},
\]%
 in terms of which the Hamiltonian $\mathcal{H}$ for the electron is%
\begin{equation}
\label{eq:H}
\mathcal{H}=\frac{1}{2m^{\ast}}\!\left(\Pi_{x}^{2}+\Pi_{y}^{2}\right)\equiv\frac{1}{2m^{\ast}}\Pi^{2}.
\end{equation}%
 After simple calculations, we obtain%
\[
dx\, d\Pi_{x}\, dy\, d\Pi_{y}=dx\, dp_{x}\, dy\, dp_{y}.
\]%
 We can then represent quantum states by quasi phase-space elements
$dxd\Pi_{x}dyd\Pi_{y}$. The Hamiltonian $\mathcal{H}$ in eq.~(\ref{eq:H})
does not depend on the position $(x,y)$. Assuming large normalization
lengths $\left(L_{1},L_{2}\right)$, we can represent the Landau states
by concentric shells of the phase space having statistical weight
$2\pi\,\Pi\,\Delta\Pi\cdot L_{1}L_{2}(2\pi\hbar)^{-2}=eBA/2\pi\hbar$,
with $A=L_{1}L_{2}$ and $\hbar\omega_\text{c}=\Delta\left(\Pi^{2}/2m^{\ast}\right)=\Pi\,\Delta\Pi/m^{\ast}$.
Hence, the LL degeneracy is given by eq.~(\ref{eq:eBAover2pihbar,A}).
Figure~\ref{cap:Landau_Quant} represents a typical Landau state in the
$\Pi_{x}$-$\Pi_{y}$ space.%

As the field $B$ is raised the separation $\hbar\omega_\text{c}$ increases,
and the quantum states are collected (or bunched) together. As a result
of bunching, the density of states $\mathcal{N}(\varepsilon)$ should
change periodically since the Landau levels are spaced equally.

The electrons obey the Fermi-Dirac statistics. Considering a system
of quasifree electrons, we define the Helmholtz free energy
$\mathcal{F}$ by%
\begin{equation}
\label{eq:F1}
\mathcal{F}=N\mu-2k_\text{B}T\cdot\sum_{i}\ln\left[1+e^{\left(\mu-E_{i}\right)/k_\text{B}T}\right],
\end{equation}%
 where $\mu$ is the chemical potential, and the factor 2 arises from spin
 degeneracy. The chemical potential $\mu$ is determined from the
 condition%
\begin{equation}
\label{eq:partialFoverpartialmu}
\frac{\partial\mathcal{F}}{\partial\mu}=0.
\end{equation}%
 The magnetization $\mathcal{M}$ for the system can be found from%
\begin{equation}
\label{eq:magnetization}
\mathcal{M}=-\frac{1}{A}\frac{\partial\mathcal{F}}{\partial
  B}.
\end{equation}%
 Equation~(\ref{eq:partialFoverpartialmu}) is equivalent to the
usual condition that the number of electrons $N$ can be obtained in
terms of the Fermi distribution function $F$:%
\begin{equation}
\label{eq:N}
N=2\sum_{i}F\!\left(E_{i}\right),\quad
F(E)\equiv\left[e^{\beta\left(E-\mu\right)}+1\right]^{-1}.
\end{equation}%
 The LL $E_{i}$ is characterized by the Landau oscillator quantum
number $\left(N_\text{L}\right)_i$.

Let us introduce the density of states $dW/dE\equiv\mathcal{N}(E)$
such that $\mathcal{N}(E)dE$ = number of states having energy
between $E$ and $E+dE$. We write eq.~(\ref{eq:F1}) in the form%
\begin{eqnarray}
\label{eq:F2}
\mathcal{F} & = & N\mu-2k_\text{B}T\!\cdot\!\int_{0}^{\infty}\!\! dE\,\frac{dW}{dE}\ln\left[1+e^{\left(\mu-E\right)/k_\text{B}T}\right]\nonumber\\
 & = & N\mu-2\int_{0}^{\infty}\!\! dE\, W(E)F(E).
\end{eqnarray}%
 The statistical weight $W$ is the total number of states having
energies less than the Landau energy
$\left(N_\text{L}+1/2\right)\hbar\omega_\text{c}$ in eq.~(\ref{eq:E,NsubL}).
This $W$ from fig.~\ref{cap:Landau_Quant} is%
\begin{equation}
\label{eq:L1L2oversquare}
W=\frac{L_{1}L_{2}}{(2\pi\hbar)^{2}}2\pi\,\Pi\,\Delta\Pi\cdot2\sum_{N_\text{L}=0}^{\infty}\!\Theta\!\left[E-\left(N_\text{L}+1/2\right)\hbar\omega_\text{c}\right],
\end{equation}%
 where $\Theta(x)$ is the Heaviside step function%
\[
\Theta(x)=\left\{ \begin{array}{cc}
0 & \textrm{if }x<0\\
1 & \textrm{if }x>0\end{array}\right..
\]%
 We introduce a dimensionless variable $\varepsilon\equiv2\pi E/\hbar\omega_\text{c}$,
and rewrite eq.~(\ref{eq:L1L2oversquare}) as%
\begin{equation}
\label{eq:WofEC}
W\!(E)=C\hbar\omega_\text{c}\!\cdot2\!\sum_{N_\text{L}=0}^{\infty}\!\Theta\!\left[\varepsilon-\left(2N_\text{L}+1\right)\pi\right],
\end{equation}%
 with $C=2\pi m^{\ast}A(2\pi\hbar)^{-2}$. We assume a high Fermi-degeneracy
such that $\mu\simeq\varepsilon_\text{F}\gg\hbar\omega_\text{c}$.
The sum in eq.~(\ref{eq:WofEC}) can be computed using Poisson's
summation formula~\cite{Courant1953}%
\begin{equation}
\label{eq:sumoff}
\sum_{n=-\infty}^{\infty}f(2\pi
n)=\frac{1}{2\pi}\!\sum_{m=-\infty}^{\infty}\int_{-\infty}^{\infty}d\tau
f(\tau)e^{-im\tau},
\end{equation}%
 where $\sum_{n=-\infty}^{\infty}f(2\pi n+t)$, $0\leq t<2\pi$ is
by assumption uniformly convergent. We write the sum in
eq.~(\ref{eq:WofEC}) as%
\begin{eqnarray}
2\sum_{n=0}^{\infty}\Theta[\varepsilon-(2n+1)\pi] & = &
\Theta(\varepsilon-\pi)+\phi(\varepsilon;0),\label{eq:2sumTheta}\\
\phi(\varepsilon;x) & \equiv &
\sum_{n=-\infty}^{\infty}\Theta(\varepsilon-\pi-2\pi|n+x|).\nonumber
\end{eqnarray}%
 Note that $\phi(\varepsilon;x)$ is periodic in $x$. After the Fourier
expansion, we set $x=0$ to obtain eq.~(\ref{eq:2sumTheta}). By
taking the real part (Re) of eq.~(\ref{eq:2sumTheta}) and using
eq.~(\ref{eq:sumoff}), we obtain%
\begin{equation}
\label{eq:Real}
\mbox{Re\,\{
  Eq.\,(\ref{eq:2sumTheta})\}}=\frac{1}{\pi}\int_{0}^{\infty}d\tau\Theta(\varepsilon-\tau)+\frac{2}{\pi}\sum_{\nu=1}^{\infty}(-1)^{\nu}\int_{0}^{\infty}d\tau\Theta(\varepsilon-\tau)\cos\nu\tau,
\end{equation}%
 where we assumed $\varepsilon\equiv2\pi E/\hbar\omega_\text{c}\gg1$, and
neglected $\pi$ against $\varepsilon$. The integral in the first
term in eq.~(\ref{eq:Real}) yields $\varepsilon$. The integral
in the second term can be evaluated by integration by parts, and using
$d\Theta/dy=\delta (y)$. We obtain%
\[
\int_{0}^{\infty}d\tau\Theta(\varepsilon-\tau)\cos\nu\tau=\frac{1}{\nu}\sin\nu\varepsilon.
\]%
 Hence,%
\begin{equation}
\label{eq:Real2}
\mbox{Re\{
  Eq.\,(\ref{eq:2sumTheta})\}}=\frac{1}{\pi}\varepsilon+\frac{2}{\pi}\sum_{\nu=1}^{\infty}\frac{(-1)^{\nu}}{\nu}\sin\nu\varepsilon.
\end{equation}%
 Using eqs.~(\ref{eq:WofEC}) and (\ref{eq:Real2}), we obtain%
\begin{equation}
\label{eq:WofE}
W\!(E)=W_{0}+W_\text{osc},
\end{equation}%
\begin{eqnarray}
W_{0} & = & C\hbar\omega_\text{c}\frac{\varepsilon}{\pi}=A\frac{m^{\ast}\!E}{\pi\hbar^{2}}\\
W_\text{osc} & = & C\hbar\omega_\text{c}\,\frac{2}{\pi}\sum_{\nu=1}^{\infty}\frac{(-1)^{\nu}}{\nu}\sin\left(\frac{2\pi\nu
  E}{\hbar\omega_\text{c}}\right).\label{eq:Wsubosc}
\end{eqnarray}%
 The oscillatory term $W_\text{osc}$ contains an infinite sum
with respect to $\nu$, but only the first term, $\nu=1$, is important
in practice as we see later. This term $W_\text{osc}$ can generate
magnetic oscillations. There is no term proportional to $B^{2}$ generating
the Landau diamagnetism. This is unexpected. We briefly discuss the
difference between 2D and 3D systems.

In 3D, the LL $E$ is given by%
\[
E=\left(N_\text{L}+\frac{1}{2}\right)\hbar\omega_\text{c}+\frac{p_{z}^{2}}{2m^{\ast}}.
\]%
 The energy $E$ is continuous in the bulk limit. The statistical
weight $W'$ is the total number of states having energies less than
$E$. The allowed values of $p_{z}$ are distributed over the range
in which $\left|p_{z}\right|$ does not exceed $\left\{2m^{\ast}\!\left[E-\left(N_\text{L}+1/2\right)\hbar\omega_\text{c}\right]\right\}^{1/2}$.
We obtain%
\begin{equation}
\label{eq:WofE2}
W'(E)=C'\frac{\left(\hbar\omega_\text{c}\right)^{3/2}}{\sqrt{2}\pi}2\!\sum_{N_\text{L}=0}^{\infty}\sqrt{\varepsilon-\left(2N_\text{L}+1\right)\pi},
\end{equation}%
 where $C'\equiv V\left(2\pi m^{\ast}\right)^{3/2}/(2\pi\hbar)^{3}$,
$\varepsilon\equiv2\pi E/\hbar\omega_\text{c}$, and $V\equiv L_{1}L_{2}L_{3}$
is the sample volume. We proceed similarly, and obtain%
\begin{equation}
\label{eq:W}
W'(E)=W'_{0}+W'_\text{L}+W'_\text{osc},
\end{equation}%
\begin{eqnarray}
W'_{0} & = & C'\frac{4}{3\sqrt{\pi}}E^{3/2}\label{eq:Wsub0p}\\
W'_\text{L} & = & -C'\frac{1}{24\sqrt{\pi}}\frac{\left(\hbar\omega_\text{c}\right)^{2}}{E^{1/2}}\label{eq:WsubLp}\\
W'_\text{osc} & = &
C'\frac{1}{\sqrt{2}}\!\left(\frac{\hbar\omega_\text{c}}{\pi}\right)^{3/2}\sum_{\nu=1}^{\infty}\frac{(-1)^{\nu}}{\nu^{3/2}}\sin\left(\frac{2\pi\nu
  E}{\hbar\omega_\text{c}}-\frac{\pi}{4}\right).\label{eq:Wsuboscp}
\end{eqnarray}%

In detail, we write the sum in eq.~(\ref{eq:WofE2}) as%
\begin{eqnarray}
2\sum_{n=0}^{\infty}\sqrt{\varepsilon-(2n+1)\pi} & = & (\varepsilon-\pi)^{1/2}+\psi(\varepsilon;0),\label{eq:2sum}\\
\psi(\varepsilon;x) & \equiv &
\sum_{n=-\infty}^{\infty}(\varepsilon-\pi-2\pi|n+x|)^{1/2}.\nonumber
\end{eqnarray}%
Since $\psi(\varepsilon;x)$ is periodic in $x$, we can use it for the
Fourier expansion of eq.~(\ref{eq:2sum}), and then set $x=0$. By taking
the real part (Re) of eq.~(\ref{eq:2sum}) and using
eq.~(\ref{eq:sumoff}), we obtain%
\begin{equation}
\mbox{Re\,\{
  Eq.\,(\ref{eq:2sum})\}}=\frac{1}{\pi}\int_{0}^{\varepsilon}d\tau(\varepsilon-\tau)^{1/2}+\frac{2}{\pi}\sum_{\nu=1}^{\infty}(-1)^{\nu}\int_{0}^{\varepsilon}d\tau(\varepsilon-\tau)^{1/2}\cos\nu\tau,\label{eq:Real3}
\end{equation}%
 where we neglected $\pi$ against $\varepsilon$. The integral in
the first term in eq.~(\ref{eq:Real3}) yields $(2/3)\varepsilon^{3/2}$,
leading to $W_{0}^{\prime}$ in eq.~(\ref{eq:Wsub0p}). The integral
in the second term can be written after integrating by part, and
changing the variable $(\nu\varepsilon-\nu\tau=t)$ as%
\[
\frac{1}{2\nu^{3/2}}\left[\sin\nu\varepsilon\int_{0}^{\nu\varepsilon}dt\frac{\cos
    t}{\sqrt{t}}-\cos\nu\varepsilon\int_{0}^{\nu\varepsilon}dt\frac{\sin
    t}{\sqrt{t}}\right].
\]%
 We now use asymptotic expansions for $\nu\varepsilon=x\gg1$:%
\begin{eqnarray*}
\int_{0}^{x}dt\frac{\sin t}{\sqrt{t}} & \sim & \sqrt{\frac{\pi}{2}}-\frac{\cos x}{\sqrt{x}}-\cdots\\
\int_{0}^{x}dt\frac{\cos t}{\sqrt{t}} & \sim &
\sqrt{\frac{\pi}{2}}+\frac{\sin x}{\sqrt{x}}-\cdots.
\end{eqnarray*}%
 The second terms in the expansion lead to $W_\text{L}^{\prime}$ in eq.~(\ref{eq:WsubLp}),
where we used the identity%
\[
\sum_{\nu=1}^{\infty}\frac{(-1)^{\nu-1}}{\nu^{2}}=\frac{\pi^{2}}{12}.
\]%
 The first terms lead to the oscillatory term $W_{\textrm{osc}}^{\prime}$
in eq.~(\ref{eq:Wsuboscp}).

The term $W_{0}^{\prime}$, which is independent of $B$, gives the
weight equal to that for a free-electron system with no field. The
term $W_\text{L}^{\prime}$ proportional to $B^{2}$ is negative (diamagnetic),
and can generate a Landau diamagnetic moment.

The energy $E$ of the 3D system is continuous, and hence the system is
manageable or soft. In contrast, the energy $E$ of the 2D system is
discrete, and hence the system is less manageable. This explains the
absence of Landau diamagnetism for the 2D system.

The statistical weight $W_\text{osc}$ in eq.~(\ref{eq:Wsubosc})
has a sine term. Hence, the density of states, $\mathcal{N}=dW/dE$,
has an oscillatory part of the form%
\[
\sin\left(\frac{2\pi E}{\hbar\omega_\text{c}}\right),\quad
E\equiv\frac{\Pi^{2}}{2m^{\ast}}.
\]%
If the density of states oscillates violently in the drop of the Fermi
distribution function: $F(E)\equiv\left[e^{\beta(E-\mu)}+1\right]^{-1}$,
the delta-function replacement formula%
\[
-\frac{dF}{dE}=\delta(E-\mu),
\]%
 cannot be used. The width of $-dF/dE$ is of the order $k_\text{B}T$.
 The critical temperature $T_\text{c}$ below which oscillations can be
 observed is $k_\text{B}T_\text{c}\sim\hbar\omega_\text{c}$. For
 $T<T_\text{c}$, we may proceed as follows. Let us consider the integral%
\[
I=\int_{0}^{\infty}\! dE\, F(E)\sin\left(2\pi
E/\hbar\omega_\text{c}\right).
\]%
 We introduce a new variable $\zeta\equiv\beta(E-\mu)$, and extend
the lower limit to $-\infty$ (low-temperature limit) so that%
\[
\int_{0}^{\infty}\!
dE\cdots\frac{1}{e^{\beta(E-\mu)}+1}=\frac{1}{\beta}\int_{-\mu\beta}^{\infty}d\zeta\cdots\frac{1}{e^{\zeta}+1}\rightarrow\frac{1}{\beta}\int_{-\infty}^{\infty}\!
d\zeta\cdots\frac{1}{e^{\zeta}+1}.
\]%
 With the help of the standard integral%
\[
\int_{-\infty}^{\infty}d\zeta\,\frac{e^{ia\zeta}}{e^{\zeta}+1}=\frac{\pi}{i\sinh\pi
  a},
\]%
 we obtain%
\begin{equation}
\label{eq:I}
I=-\frac{\pi
  k_\text{B}T\cos\left(2\pi\varepsilon_\text{F}/\hbar\omega_\text{c}\right)}{\sinh\left(2\pi^{2}k_\text{B}Tm^{\ast}/\hbar eB\right)}.
\end{equation}%
 For very low fields the oscillation number in the range $k_\text{B}T$
becomes great, and hence the sinusoidal contribution must cancel out.
This effect is represented by the factor%
\[
\left[\sinh\left(2\pi^{2}k_\text{B}Tm^{\ast}/\hbar eB\right)\right]^{-1}.
\]

We calculate the free energy indicated in eq.~(\ref{eq:F2}) using the
statistical weight $W$ in eq.~(\ref{eq:WofE}), and obtain%
\begin{equation}
\mathcal{F}=N\mu-2A\frac{m^{\ast}}{\pi\hbar^{2}}\varepsilon_\text{F}+2A\frac{eB}{\pi\hbar}
k_\text{B}T\cdot\sum_{\nu=1}^{\infty}\frac{(-1)^{\nu}}{\nu}\frac{\cos\left(2\pi\nu\varepsilon_\text{F}/\hbar\omega_\text{c}\right)}{\sinh\left(2\pi^{2}\nu
  k_\text{B}Tm^{\ast}/\hbar eB\right)},\label{eq:F3}
\end{equation}%
 where we used the integration formula~(\ref{eq:I}), and took the
 low-temperature limit except for the oscillatory terms. The
magnetization $\mathcal{M}$ can be obtained using eq.~(\ref{eq:magnetization}).

So far, we have not considered the Pauli spin magnetization~\cite{Pauli1927}%
\[
\mathcal{M}_\text{Pauli}=2\mu_\text{B}^{2}B\mathcal{N}_{0}\!\left(\varepsilon_\text{F}\right)\!/A=2n\mu_\text{B}^{2}B/\varepsilon_\text{F},
\]%
 with $\mu_\text{B}\textrm{ (Bohr magneton)}=e\hbar/2m$ and $n
\textrm{ (electron number density)}=m^{\ast}\varepsilon_\text{F}/\pi\hbar^{2}$. Using
eqs.~(\ref{eq:magnetization}), (\ref{eq:N}) and (\ref{eq:F3}), we obtain
the total magnetization
$\mathcal{M}_\text{tot}=\mathcal{M}_\text{Pauli}+\mathcal{M}_\text{osc}$:%
\begin{equation}
\label{eq:Mtot}
\mathcal{M}_\text{tot}=2n\mu_\text{B}\frac{\mu_\text{B}B}{\varepsilon_\text{F}}\left[1+2\frac{k_\text{B}T}{\mu_\text{B}B}\frac{m}{m^{\ast}}\frac{\cos\left(2\pi\varepsilon_\text{F}/\hbar\omega_\text{c}\right)}{\sinh\left(2\pi^{2}k_\text{B}T
 m^{\ast}/\hbar eB\right)}\right].
\end{equation}%
The $B$-dependence of $\mathcal{F}$ is contained in the last term in
eq.~(\ref{eq:F3}). The linear $B$-dependence of the multiplication
factor is much stronger than the $B$-dependence of the alternating series.
Therefore, the contribution from the $B$-derivative of the series is
neglected. The variation of the statistical weight $W$ is periodic in
$B^{-1}$, but it is far from sinusoidal. Only the first oscillatory
term, $\nu=1$, is important and kept in eq.~(\ref{eq:Mtot}) since
$\sinh\left(2\pi^{2}k_\text{B}Tm^{\ast}/\hbar eB\right)\gg1$. The width
of $dF/dE$ is finite for a finite $T$. In this $E$-range, many
oscillations can occur if the field $B$ is made low. We assumed this
condition to obtain eq.~(\ref{eq:Mtot}). The magnetic susceptibility
$\chi$ is defined by the ratio $\chi=\mathcal{M}/B$.

In conclusion, the 2D system is intrinsically paramagnetic since the Landau's diamagnetic
term is absent, but the system exhibits a dHvA oscillation.

\end{document}